# Image Enhancement via Bilateral Learning

Saeedeh Rezaee, Nezam Mahdavi-Amiri
Faculty of Mathematical Sciences
Sharif University of Technology
Tehran, Iran

*Abstract*— Nowadays, due to advanced digital imaging technologies and internet accessibility to the public, the number of generated digital images has increased dramatically. Thus, the need for automatic image enhancement techniques is quite apparent. In recent years, deep learning has been used effectively. Here, after introducing some recently developed works on image enhancement, an image enhancement system based on convolutional neural networks is presented. Our goal is to make an effective use of two available approaches, convolutional neural network and bilateral grid. In our approach, we increase the training data and the model dimensions and propose a variable rate during the training process. The enhancement results produced by our proposed method, while incorporating 5 different experts, show both quantitative and qualitative improvements as compared to other available methods.

## I. Introduction

The accessibility of camera phones and the emergence of various online applications for image sharing have led to the rapid expansion of available image data. Due to the sheer number of digital images taken by unskilled ordinary photographers, the need for effective tools, or automatic image enhancement algorithms, to enhance the quality of these images is ever more apparent. Image enhancement, as an important field in digital image processing with widespread applications, has been extensively studied for several decades. Image enhancement is even more important for camera phones, since most images taken with camera phones have low quality, with defects such as color imbalance or low contrast. Professional image restoration can significantly reduce or eliminate these defects, albeit is expensive due to manual, tedious and time-consuming process. It is particularly challenging, since even the simple contrast and brightness changes depend on the content of the image under consideration, and are subject to personal judgment.

Deep learning has recently been used for image enhancement. Traditionally introduced for computer vision applications, deep learning is associated with various model architectures using basic building blocks, one of which being convolutional neural networks (CNNs). CNNs are designed to extract optimal features hierarchically from an input image. These features, in turn, can be used to learn the transformation applied by professional photo retouching. These photo retouchings, collected from 5 individual professional photographers, enhance a fixed set of images that form the training data for deep learning. Since the objective of such a deep model is to enhance amateur images, we adopt the model architecture of [1] applying a bilateral grid as the last layer of the network. Here, we introduce three main novelties having significant impacts on the automatic image enhancement for amateur photography: (1) instead of fitting a specific model for each professional retoucher, we introduce a unified model, making the subsequent selection of the individual models irrelevant, (2) we increase the size of the model to increase the learning capacity of the unified model, and (3) we adopt a dynamic learning rate schedule to improve the quality of the training.

## II. Related work

Filtering is used to enhance images. Bilateral filtering has been commonly used for this purpose, which, in addition to reducing the noise and smoothing the image, can also preserve the edges [2]. Other filtering methods such as high-dimensional Gaussians can introduce a way to increase the speed of bilateral filtering to filter images [3], while joint bilateral upsampling can use the bilateral filtering to produce high resolution [4]. The guided filter is an edge smoothing operator, which behaves better near the edges [5, 6]. Though joint bilateral upsampling and guided filtering in general can be fast enough to work on the computationally-limited mobile environments, they are limited to a range of simple operators. In [7], a generalized bilateral guided upsampling method was developed, while the authors of [8] proposed a method to reconstruct the depth map. The authors of [9] proposed a deep network using two separate stacks, one making an inaccurate global prediction based on the whole content of the input image, and the other extending the prediction locally. Deep learning was also used for image super-resolution [10], where a direct end-to-end mapping between low-resolution and high-resolution images is being learned. Deep neural networks were also used for denoising [11, 12], and automatic image matting for portrait images using CNNs [13]. Conditional adversarial networks can particularly be used for image-to-image transformation applications, where the model learns the mapping from the input to the output image [14]. In [15], a deep network consisting of several recurrent layers and CNNs were used for image enhancement.

In other studies, automatic image enhancement was tailored to focus on user preferences learned from the training set [16]. In [17], 5000 images were collected as the training data, with 5 professional photographers being asked to retouch the images. The authors of [18] considered finding the best transformation between the input and the output for each pixel of an image using a coarse-to-fine color manipulation. This method estimates the transformation to the entire image considering the local context. In contrast, the authors of [19] took the account of both local and global image contents, and the authors of [20] developed descriptors for image enhancement. Here, we propose an extension of the deep bilateral grid processing, while making use of both local and global features to directly learn the end-to-end image transformation.

### III. ARCHITECTURE

The proposed architecture consists of two separate paths that are merged into a layer called fusion before final prediction. The network learns to generate a full resolution output using the grayscale map and the slicing layer. The architecture of this network is shown in Figure 1.

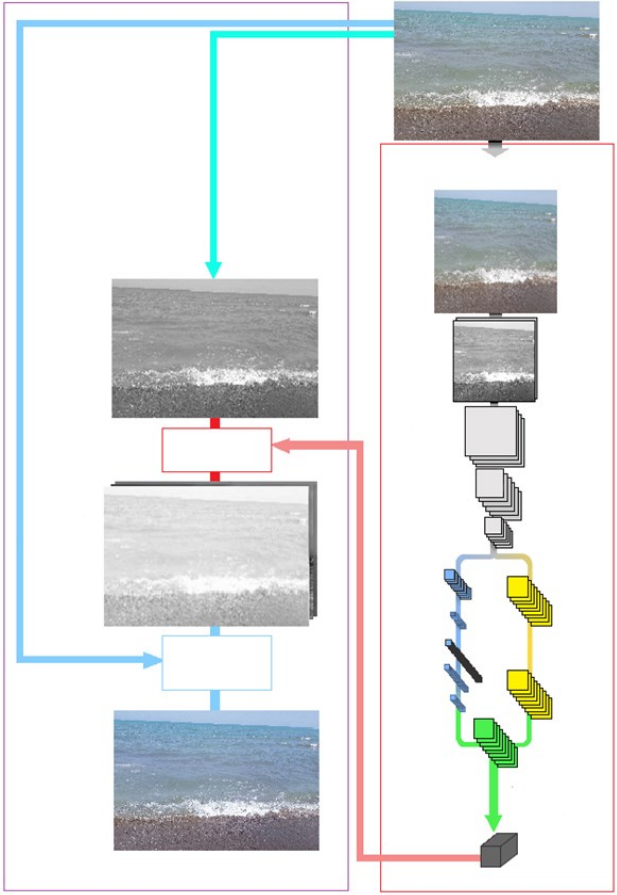

Fig. 1. Network architecture.

### A. Low-resolution Stream

*1) Low-level features path:* The Input $\tilde{I}$ is processed by a stack of convolutional layers $(S^i)_{i=1,\ldots,n_S}$ to extract low level properties and reduce spatial resolution. The size of the input $\tilde{I}$, in the low resolution branch is 256×256. First, the low-resolution image $S^0 := \tilde{I}$, is processed with a stack of standard convolutional layers with stride $s = 2$:

$$S_c^i[x,y] = \qquad (1)$$

$$\sigma\left(b_c^i + \sum_{x',y',c'} w_{cc'}^i[x',y'] S_{c'}^{i-1}[sx+x', sy+y']\right),$$

where $\sigma(\cdot) = \max(0,\cdot)$, $i = 1,\ldots,n_s$ corresponds to each layer, $c$ and $c'$ are the channels within layers, $w^i$ is the array of weights for convolution layers, $b^i$ is the vector of bias, and $x'$ and $y'$ are 3×3 convolutions. The ReLU activation function and zero-padding are being used in all convolutions.

*2) Local path:* In the local path, the last layer of low-resolution stream, $L^0 := S^{n_S}$, is processed by a stack of convolutional layers as

$$L_c^i[x,y] = \qquad (2)$$

$$\sigma\left(b_c^i + \sum_{x',y',c'} w_{cc'}^i[x',y'] L_{c'}^{i-1}[sx+x', sy+y']\right),$$

where $i = 1,\ldots,n_L$ and $s = 1$.

*3) Global path:* The global path begins with $G^0 := S^{n_S}$ which, like the local path, includes two strided convolutional layers with $s = 2$:

$$G_c^i[x,y] = \qquad (3)$$

$$\sigma\left(b_c^i + \sum_{x',y',c'} w_{cc'}^i[x',y'] G_{c'}^{i-1}[sx+x', sy+y']\right).$$

These layers are followed by three fully-connected layers, defined by

$$G_c^i[x,y] = \sigma\left(\sum_{c'} w_{cc'}^i G_{c'}^{i-1}[x,y] + b_c^i\right), i = 3,4,5 \qquad (4)$$

*4) Fusion layer:* The local and global paths are fused with a pointwise affine transformation as follows:

$$F_c[x,y] = \sigma\left(b_c + \sum_{c'} w'_{cc'} G_{c'}^{n_G} + \sum_{c'} w_{cc'} L_{c'}^{n_L}[x,y]\right), \qquad (5)$$

where $G_{c'}^{n_G}$ and $L_{c'}^{n_L}$ are the results of local and global feature paths, yielding an array of features having the size 16×16×64. Finally, a linear prediction filter with the dimension 64×1×1 is applied to produce a 16×16 map for 96 channels as

$$A_c[x,y] = b_c + \sum_{c'} F_{c'}[x,y] w_{cc'}. \qquad (6)$$

### B. Bilateral Grid

We have focused on the neural network architecture of the model so far. Now, we define the final feature map $A$ as a bilateral grid. To facilitate this, the final feature map $A$ is defined as a bilateral grid using a partial notation:

$$A_{dc+z}[x,y] \leftrightarrow A_c[x,y,z] \,, \tag{7}$$

where $A$ is a bilateral grid of size 16×16×8, $d$ is the depth of 8 and each cell contains 12 values.

*C. Slicing Layer*

Slicing layer takes single-channel grayscale map $g$ and bilateral feature map $A$, having a much lower spatial resolution than $g$, as an input and generates new feature map $\bar{A}$ as an output. The new feature map is calculated by tri-linearly interpolation as follows:

$$\bar{A}_c[x,y] = \tag{8}$$

$$\sum_{i,j,k} \tau(s_x x - i)\tau(s_y y - j)\tau(d.g[x,y] - k)A_c[i,j,k] \,,$$

where the linear interpolation kernel is defined as $\tau(\cdot) = \max(1 - |\cdot|, 0)$ and $s_x$ and $s_y$ are the width and height ratios of the grid's dimensions with respect to the full-resolution image dimensions. Basically, a vector of coefficients is assigned to each pixel such that the depth in the grid is given by the gray scale value $g[x,y]$. The spatial resolution of the grid is fixed to be 16×16, and $d$ is the depth of grid with the size of 8.

*D. Full-resolution Stream*

*1) Grayscale map:* Grayscale guidance map $g$ is defined as a pointwise nonlinear transformation of the full-resolution features:

$$g[x,y] = b + \sum_{c=0}^{2} \rho_c(M_c^T.\phi_c[x,y] + b'_c) \,, \tag{9}$$

$$\rho_c(x) = \sum_{i=0}^{15} a_{c,i} \max(x - t_{c,i}, 0) \,, \tag{10}$$

where $M_c^T$ is the row of a 3×3 color transformation matrix, $\phi$ is the input $I$, $b$ and $b'_c$ are scalar biases, and $\rho_c$ is a piecewise linear function parametrized as the sum of 16 ReLU functions. ReLU functions are defined by $t_{c,i}$ and $a_{c,i}$ as threshold and slopes respectively. We initialize $M$ to be identity matrix and $a, b, b'$ and $t$ such that $\rho_c$ is an identity mapping over [0, 1] and they are learned jointly along with network parameters.

*2) Final output:* Each channel of the final output $O_c$ is defined by the channels of the sliced feature map $\bar{A}$ as

$$O_c[x,y] = \tag{11}$$

$$\bar{A}_{n_\phi + (n_\phi+1)c} + \sum_{c'=0}^{n_\phi - 1} \bar{A}_{c' + (n_\phi+1)c}[x,y]\phi_{c'}[x,y] \,,$$

where $c$ is the index of the final output channel, $\phi$ is the full-resolution input and $n_\phi = 3$.

*E. Training*

The network is trained with the dataset of full-resolution input-output pairs for a given operator. The weights of layers are optimized by minimizing the $L_2$ norm of the loss function on the training dataset. Batch normalization is used and network parameters are optimized by the Adam algorithm with a learning rate of $10^{-4}$.

IV. OUR METHOD

Our proposed method includes three strategies to improve the performance of the network. The first strategy is to increase the amount of the data by combining 5 different experts. We also increase learning capacity by adding the dimensions of the model. Finally, we gradually decrease the learning rate during training, as describe below.

*A. Increased Data*

Here, we use the MIT database "FiveK" collected by Bychkovsky et al. [17]. The data set is 5,000 images retouched by 5 experts (a total of 25,000 images) and each time the model is trained according to the editions of one of the experts. In many cases, the results of a model trained by retouching images of one specialist might not be as high quality as the results obtained from a model trained with the images edited by other experts. Besides, it is difficult for the user to compare and select the final improved image from these different outputs. Therefore, we expand the data in order to resolve these issued. The model is trained simultaneously by 5 different experts, so that the decision can be made automatically by the model and eventually a final result is visible to the user.

*B. Increasing the Dimensions of the Model*

By increasing the data, a much larger model is required. Table 1 summarizes the dimensions of each layer in our proposed arcitecture.

TABALE. 1. Dimensions of the network architecture. $f$, $fc$ define fusion, fully connected networks, and $c$ and $l$ define convolutional network and pointwise linear transformation.

|  | Type | Size | Channels |
|---|---|---|---|
| $S^1$ | c | 128 | 8 |
| $S^2$ | c | 64 | 16 |
| $S^3$ | c | 32 | 32 |
| $S^4$ | c | 16 | 64 |
| $L^1$ | c | 16 | 128 |
| $L^2$ | c | 16 | 128 |
| $G^1$ | c | 8 | 128 |
| $G^2$ | c | 4 | 128 |
| $G^3$ | fc | - | 512 |
| $G^4$ | fc | - | 256 |
| $G^5$ | fc | - | 128 |
| F | f | 16 | 128 |
| A | l | 16 | 96 |

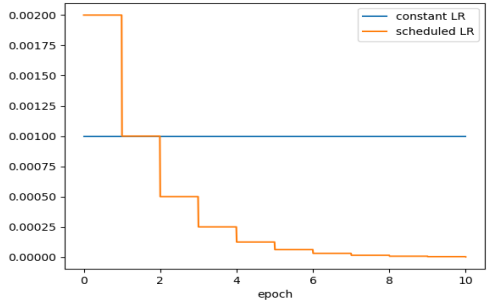

Fig. 2. Learning rate during training.

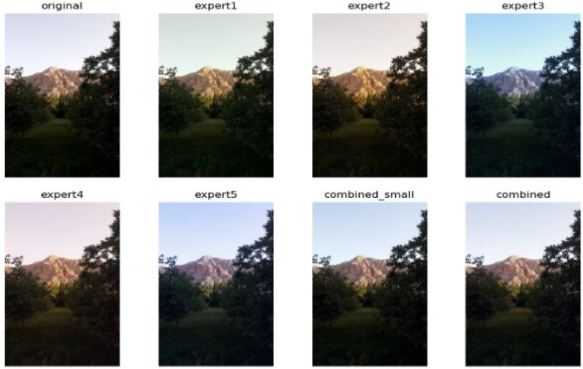

Fig. 3. Comparing output results: results of models trained by 5 different experts vs. the combined model.

*C. Learning Rate*

We use a gradually decreasing learning rate in contrast to a fixed learning rate for optimization. Figure 2 shows the learning rate changes per epoch.

*D. Results*

The model is trained with a batch size of 4 and proposed learning rates. The model is trained in 10 epochs, typically lasting one day using the Nvidia 1060 Ti GPU.

We compare the performances of the proposed model and 5 different models trained with one of the expert inputs. We also trained the original model with the combined data named as combines_small model. We used random images taken from different parts of Iran by mobile phone without any retouching to test the model. As shown in Figure 3, the result of the proposed model (combined model) has higher quality than the outputs of the individual model trained on each expert and the original model trained on combined data (combined_small).

In addition to visual comparison of the results, numerical comparison of the loss functions is illustrated in Figure 4, where the loss values are stored randomly during the training. It is shown that the loss value of the proposed model with dynamic learning rate (combined_lrs) is smaller than the loss values of the original model and the proposed model with a fixed learning rate.

We have also used PSNR as another criterion for evaluating the network, where the values are stored randomly during network training. The results shown in Figure 5 suggest that the maximum PSNR value is obtained by the combined model with dynamic learning rate.

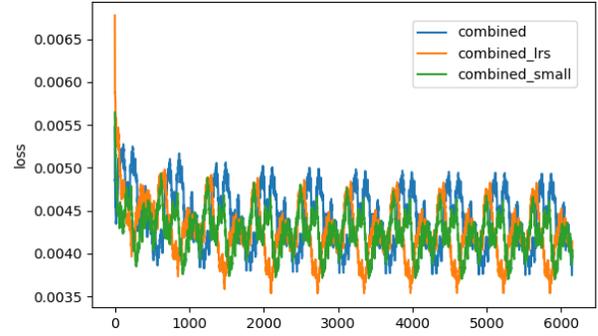

Fig. 4. Comparison of loss function. The blue plot shows the results of the combined model with fixed learning rate, the orange plot shows the loss of the proposed model with dynamic learning rate and the green line shows the loss function of the original model with fixed learning rate.

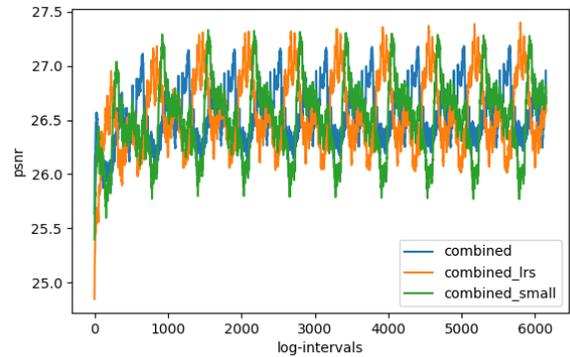

Fig. 5. Comparison of PSNR of different models. The blue line shows the results of the combined model with fixed learning rate, the orange line shows the PSNR of the proposed model with dynamic learning rate and the green line shows the results of the original model.

## V. CONCLUSION

We adopted a combined convolutional neural network along with a bilateral grid to improve the image quality. Also, a slicing layer was applied for upsampling in the model architecture. We proposed a method generalizing the learning process by combining the data from different experts. The model observes the images retouched by 5 different experts in the training set and is used to improve any random image automatically. We also increased the dimensions of the model and applied a dynamic learning rate to improve the performance of the model. Expanding the dimensions of the model and gradually reducing the learning rate lead to increasing the speed and improving the learning quality. The implementation results on a new data set affirmed a good performance of the proposed method.